\documentclass[pre,preprint,showpacs,showkeys,preprintnumbers,amsmath,amssymb]{revtex4}
\usepackage{graphicx}
\usepackage{dcolumn}
\usepackage{bm}
\begin{document}
\title{Synthetic Turbulence, Fractal Interpolation and Large-Eddy Simulation}
\author{Sukanta Basu}
\email{basus@msi.umn.edu}
\affiliation{St. Anthony Falls Laboratory, University of Minnesota, Minneapolis, MN 55414}
\author{Efi Foufoula-Georgiou}
\altaffiliation[Also at ]{the National Centre for Earth Surface Dynamics.}
\author{Fernando Port\'{e}-Agel}
\altaffiliation[Also at ]{the National Centre for Earth Surface Dynamics.}
\affiliation{St. Anthony Falls Laboratory, University of Minnesota, Minneapolis, MN 55414}
\date{\today}

\begin{abstract}
Fractal Interpolation has been proposed in the literature as an efficient way to construct closure models for the 
numerical solution of coarse-grained Navier-Stokes equations. It is based on synthetically generating a scale-invariant 
subgrid-scale field and analytically evaluating its effects on large resolved scales. In this paper, we propose an 
extension of previous work by developing a multiaffine fractal interpolation scheme and demonstrate that it preserves 
not only the fractal dimension but also the higher-order structure functions and the non-Gaussian probability density 
function of the velocity increments. Extensive a-priori analyses of atmospheric boundary layer measurements further 
reveal that this Multiaffine closure model has the potential for satisfactory performance in large-eddy simulations. 
The pertinence of this newly proposed methodology in the case of passive scalars is also discussed.
\end{abstract}


\maketitle
\medskip
\section{\label{Sec1}Introduction} 

Generation of turbulence-like fields (also known as \textit{Synthetic Turbulence}) has received considerable attention 
in recent years. Several schemes have been proposed \cite{Vicsek,Benzi,Juneja,Biferale,Bohr} with different degrees of 
success in reproducing various characteristics of turbulence. Recently, Scotti and Meneveau \cite{Scotti97,Scotti99} 
further broadened the scope of synthetic turbulence research by demonstrating its potential in computational modeling. 
Their innovative turbulence emulation scheme based on the \textit{Fractal Interpolation Technique} (FIT) \cite{Barn86,
Barn93} was found to be particularly amenable for a specific type of turbulence modeling, known as Large-Eddy 
Simulation (LES, at present the most efficient technique available for high Reynolds number flow simulations, in which 
the larger scales of motion are resolved explicitly and the smaller ones are modeled). The underlying idea was to 
explicitly reconstruct the subgrid (unresolved) scales from given resolved scale values (assuming computation grid-size 
falls in the \textit{inertial range} of turbulence) using FIT and subsequently estimate the relevant subgrid-scale (SGS)
tensors necessary for LES. Simplicity, straightforward extensibility for multi-dimensional cases, and low 
computational complexity (appropriate use of \textit{Fractal Calculus} can even eliminate the computationally expensive 
explicit reconstruction step, see Section \ref{Sec4} for details) makes this FIT-based approach an attractive 
candidate for SGS modeling in LES.

Although the approach of \cite{Scotti97,Scotti99} is better suited for LES than any other similar scheme (e.g., \cite{
Vicsek,Benzi,Juneja,Biferale,Bohr}), it falls short in preserving the essential small-scale properties of turbulence, 
such as multiaffinity (will be defined shortly) and non-Gaussian characteristics of the probability density function 
(pdf) of velocity increments. It is the purpose of this work to extend the approach of \cite{Scotti97,Scotti99} in 
terms of realistic turbulence-like signal generation with all the aforementioned desirable characteristics and demonstrate
its potential for LES through a-priori analysis (an LES-SGS model evaluation framework). We will also demonstrate the 
competence of our scheme in the emulation of passive-scalar fields for which the non-Gaussian pdf and multiaffinity are 
significantly pronounced and cannot be ignored.
\medskip
\section{\label{Sec2}Basics of Fractal Interpolation} 

The fractal interpolation technique is an iterative affine mapping procedure to construct a synthetic deterministic 
small-scale field (in general fractal provided certain conditions are met, see below) given a few large-scale 
interpolating points (anchor points). For an excellent treatise on this subject, the reader is referred to the book by 
Barnsley \cite{Barn93}. In this paper, we will limit our discussion (without loss of generality) only to the case of 
three interpolating data points: $\left\{ {\left( {x_i ,\tilde {u}_i } \right),\mbox{ }i = 0,1,2} \right\}$. For this case, the 
fractal interpolation iterative function system (IFS) is of the form $\left\{ {{\rm R}^2;w_n ,n = 1,2} \right\}$, where,
$w_n $ have the following affine transformation structure:
\begin{equation}\label{eq1}
w_n \left( {{\begin{array}{*{20}c}
 x \hfill \\
 u \hfill \\
\end{array} }} \right) = \left[ {{\begin{array}{*{20}c}
 {a_n } \hfill & 0 \hfill \\
 {c_n } \hfill & {d_n } \hfill \\
\end{array} }} \right]\left( {{\begin{array}{*{20}c}
 x \hfill \\
 u \hfill \\
\end{array} }} \right) + \left( {{\begin{array}{*{20}c}
 {e_n } \hfill \\
 {f_n } \hfill \\
\end{array} }} \right),\mbox{ }n = 1,2.
\end{equation}
To ensure continuity, the transformations are constrained by the given data points as follows:
$w_n \left( {{\begin{array}{*{20}c}
 {x_0 } \hfill \\
 {\tilde {u}_0 } \hfill \\
\end{array} }} \right) = \left( {{\begin{array}{*{20}c}
 {x_{n - 1} } \hfill \\
 {\tilde {u}_{n - 1} } \hfill \\
\end{array} }} \right)$ and $w_n \left( {{\begin{array}{*{20}c}
 {x_2 } \hfill \\
 {\tilde {u}_2 } \hfill \\
\end{array} }} \right) = \left( {{\begin{array}{*{20}c}
 {x_n } \hfill \\
 {\tilde {u}_n } \hfill \\
\end{array} }} \right),$ for $\mbox{ }n = 1,2.$
The parameters $a_n ,c_n ,e_n \mbox{ and }f_n $ can be easily determined in terms of $d_n $ (known as the vertical 
stretching factors) and the given anchor points $\left( {x_i ,\tilde {u}_i } \right)$ by solving a linear system of equations. 
The attractor of the above IFS, $G$, is the graph of a continuous function $u:[x_0 ,x_2 ] \to {\rm R}$, which 
interpolates the data points $\left( {x_i ,\tilde {u}_i } \right)$, provided the vertical stretching factors $d_n $ obey $0 \le 
\left| {d_n } \right| < 1$. In other words,
\begin{equation}\label{eq2}
\begin{array}{l}
 G = \left\{ {\left( {x,u\left( x \right)} \right):x \in \left[ {x_0 ,x_2 } 
\right]} \right\}, \\ 
 \mbox{where,} \\ 
 u\left( {x_i } \right) = \tilde {u}_i ,\mbox{ }i = 0,1,2. \\ 
 \end{array}
\end{equation}
Moreover, if $\left| {d_1 } \right| + \left| {d_2 } \right| > 1$ and $\left( {x_i ,\tilde {u}_i } \right)$ are not collinear, 
then the fractal (box-counting) dimension of $G$ is the unique real solution $D$ of $\left| {d_1 } \right|a_1^{D - 1} + 
\left| {d_2 } \right|a_2^{D - 1} = 1$ (for rigorous proof see \cite{Barn86}). In the special case of three equally 
spaced points covering the unit interval [0,1], i.e., $x_0 = 0,\mbox{ }x_1 = 0.5\mbox{ and }x_2 = 1$, the parameters of 
the affine transformation kernel become: $a_n = 0.5;\mbox{ }c_n = \left( {\tilde {u}_n - \tilde {u}_{n - 1} } \right) - 
d_n \left( {\tilde {u}_2 - \tilde {u}_0 } \right);\mbox{ }e_n = x_{n - 1} ;\mbox{ }f_n = \tilde {u}_{n - 1} - d_n \tilde {u}_0 ;
\mbox{ }n = 1,2.$ In this case, the solution for the fractal dimension $\left( D \right)$ becomes:
\begin{equation}\label{eq3}
D = 1 + \log_2 \left( {\left| {d_1 } \right| + \left| {d_2 } \right|} 
\right)
\end{equation}
Notice that the scalings $d_1 $ and $d_2 $ are free parameters and cannot be determined using only equation (\ref{eq3});
at least one more constraint is necessary. For example, \cite{Scotti97,Scotti99} chose to use the additional 
condition: $\left| {d_1 } \right| = \left| {d_2 } \right|$.
\medskip
\section{\label{Sec3}Synthetic Turbulence Generation} 

Not long ago, it was found that turbulent velocity signals at high Reynolds numbers have a fractal dimension of $D 
\simeq 1.7\pm 0.05$, very close to the value of $D = 5/3$ expected for Gaussian processes with a $ -5/3$
spectral slope \cite{Scotti95}.  For $D = 5/3$, the assumption of $\left| {d_1 } \right| = \left| {d_2 } 
\right|$ along with Equation \ref{eq3} yields $\left| {d_1 } \right| = \left| {d_2 } \right| = 2^{ - 1 \mathord{\left
/ {\vphantom {1 3}} \right. \kern-\nulldelimiterspace} 3}$ \cite{Scotti97,Scotti99}. One contribution of this paper is 
a robust way of estimating the stretching parameters without any ad-hoc prescription; the resulting synthetic field 
will not only preserve the fractal dimension $\left( D \right)$ but also other fundamental properties of real 
turbulence. 

As an exploratory example, using the fractal interpolation IFS (Equation \ref{eq1}), we construct a $2^{17}$ points 
long synthetic fractal series, $u\left( x \right)$, with given coarse-grained points $\left( {0.0,1.2} \right), \left( 
{0.5,-0.3} \right)\mbox{ and }\left( {1.0,0.7} \right)$ and the stretching parameters used in \cite{Scotti97,Scotti99}: 
$d_1 = - 2^{-1 \mathord{\left/ {\vphantom {1 3}} \right. \kern-\nulldelimiterspace} 3},d_2 = 2^{-1 \mathord{\left/ 
{\vphantom {1 3}} \right. \kern-\nulldelimiterspace} 3}$. Clearly, Figure 1a depicts that the synthetic series has 
fluctuations at all scales and it passes through all three interpolating points. 

Next, from this synthetic series we compute higher-order structure functions (see Figure 1b for orders $2, 4$ and $6$), 
where the $q^{th}$-order structure function, $S_q\left( r \right)$, is defined as follows:
\begin{equation}\label{eq4}
S_{q}\left( r \right) = \left\langle {\left| {u\left( {x + r} \right) - u\left( x \right)} 
\right|^q} \right\rangle \sim r^{\zeta _q }
\end{equation}
\noindent
where, the angular bracket denotes spatial averaging and $r$ is a separation distance that varies in an appropriate scaling region 
(known as the inertial range in turbulence). If the scaling exponent $\zeta _q$ is a nonlinear function of $q$, then 
following the convention of \cite{Vicsek,Benzi,Juneja,Biferale,Bohr}, the field is called \textit{multiaffine}, 
otherwise it is termed as \textit{monoaffine}. In this context, we would like to mention that, Kolmogorov's celebrated 
1941 hypothesis (a.k.a K41) based on the assumption of global scale invariance in the inertial range predicts that the 
structure functions of order $q$ scale with an exponent $q/3$ over inertial range separations \cite{Ann,Frisch}.
Deviations from $\zeta_q = q/3$ would suggest inertial range intermittency and invalidate the K41 hypothesis.     
Inertial range intermittency is still an unresolved issue, although experimental evidence for its existence is 
overwhelming \cite{Ann,Anselmet}. To interpret the curvilinear behavior of the $\zeta_q$ function observed in 
experimental measurements (e.g., \cite{Anselmet}), Parisi and Frisch \cite{Parisi,Frisch} proposed the \textit {multifractal} 
model, by replacing the global scale invariance with the assumption of local scale invariance. They 
conjectured that at very high Reynolds number, turbulent flows have singularities (almost) everywhere and showed that 
the singularity spectrum is related to the structure function-based scaling exponents, $\zeta_q$ by the Legendre 
transformation.   

Our numerical experiment with the stretching parameters of \cite{Scotti97,Scotti99}, i.e., $\left| d_1 \right| 
= \left| d_2 \right| = 2^{ - 1 \mathord{\left/ {\vphantom {1 3}} \right. \kern-\nulldelimiterspace} 3}$, revealed that 
the scaling exponents follow the K41 predictions (after ensemble averaging over one hundred realizations corresponding 
to different initial interpolating points), i.e., $\zeta_q = q/3$ (not shown here), a signature of monoaffine 
fields. Later on, we will give analytical proof that indeed this is the case for $\left| {d_1 } \right| = \left| {d_2 } 
\right| = 2^{ - 1 \mathord{\left/ {\vphantom {1 3}} \right. \kern-\nulldelimiterspace} 3}$. Also, in this case, the 
pdfs of the velocity increments, $\delta u_r \left( x \right) = u\left( {x + r} \right) - u\left( x \right)$, always 
portray near-Gaussian (slightly platykurtic) behavior irrespective of $r$ (see Figure 1c).
This is contrary to the observations \cite{Ann,Anselmet}, where, typically the pdfs of increments are found to be $r$
dependent and become more and more non-Gaussian as $r$ decreases. Theoretically, non-Gaussian characteristics of pdfs 
correspond to the presence of intermittency in the velocity increments and gradients (hence in the energy dissipation) 
\cite{Frisch,Ann,Benzi,Bohr}.

In Figure 2, we plot the wavelet spectrum of this synthetic series. Due to the dyadic nature of the fractal 
interpolation technique, the Fourier spectrum will exhibit periodic modulation (see Figures 7 and 8 of \cite{Scotti99}).
To circumvent this issue we make use of the (dyadic) discrete Haar wavelet transform. Following \cite{Katul}, the 
wavelet power spectral density function $E(K_m)$ is defined as:
\begin{equation}\label{eq5}
E\left( {K_m } \right) = \frac{\left\langle {\left( {WT^{\left( m 
\right)}\left[ i \right]} \right)^2} \right\rangle dx}{2\pi \ln \left( 2 \right)} \\
\end{equation}
where, wavenumber $K_m (= \frac{2\pi }{2^mdx})$ corresponds to scale $R_m (= 2^mdx)$. The scale index $m$ 
runs from $1$ (finest scale) to $log_2(N)$ (coarsest scale). $WT^{\left( m \right)}\left[ i \right]$, $dx$ and $N$ 
denote the Haar wavelet coefficient at scale $m$ and location $i$, spacing in physical space and length of the spatial series, 
respectively. The power spectrum displays the inertial range slope of $-5/3$, as anticipated.


At this point, we would like to invoke an interesting mathematical result regarding the scaling exponent spectrum, $
\zeta_q$, of the fractal interpolation IFS \cite{Levy}:
\begin{equation}\label{eq6}
\zeta_q = 1 - \log _N \sum\limits_{n = 1}^N {\left| {d_n } \right|^q} 
\end{equation}
where, $N$ = the number of anchor points $- 1$ (in our case $N = 2$).
The original formulation of \cite{Levy} was in terms of a more general scaling exponent spectrum, $\tau \left(q\right)$,
rather than the structure function based spectrum $\zeta_q$. The $\tau \left(q\right)$ spectrum is an exact Legendre 
tranform of the singularity spectrum in the sense that it is valid for any order of moments (including negative) and 
any singularities \cite{Muzy,Jaff}. $\tau \left(q\right)$ can be reliably estimated from data by the Wavelet-Transform 
Modulus-Maxima method \cite{Muzy,Jaff}. To derive Equation \ref{eq6} from the original formulation, we made use of 
the equality: $\tau \left( q \right) = \zeta _q - 1$, which holds for positive $q$ and for positive singularities of 
H\"{o}lder exponents less than unity \cite{Muzy,Jaff}. In turbulence, the most probable H\"{o}lder exponent is 0.33 
(corresponding to the K41 value) and for all practical purposes the values of H\"{o}lder exponents lie between $0$ and $
1$ (see \cite{Bacry,Verg}). Hence the use of the above equality is well justified.   

Equation \ref{eq6} could be used to validate our previous claim, that the parameters of \cite{Scotti97,Scotti99} 
give rise to a monoaffine field (i.e., $\zeta_q$ is a linear function of $q$). If we consider $\left| {d_1 } \right
| = \left| {d_2 } \right| = d = 2^{ - 1 \mathord{\left/ {\vphantom {1 3}} \right. \kern-\nulldelimiterspace} 3}$, then, 
$\zeta _q = 1 - \log _2 \left( {\left| {d_1 } \right|^q + \left| {d_2 } \right|^q} \right) = 1 - \log _2 \left( {2d^q} 
\right) = - q\log _2 \left( d \right) = - q\log _2 \left( {2^{ - 1 \mathord{\left/ {\vphantom {1 3}} \right. 
\kern-\nulldelimiterspace} 3}} \right) = q/3$ [QED]. Equation \ref{eq6} could also be used to derive the classic 
result of Barnsley regarding the fractal dimension of IFS. It is well-known \cite{Mando,Davis} that the graph dimension 
(or box-counting dimension) is related to $\zeta _1 $ as follows: $D = 2 - \zeta _1.$ Now, using Equation \ref{eq6} 
we get, $D = 2 - \zeta _1 = 1 + \log _N \sum\limits_{n = 1}^N {\left| {d_n } \right|}$. For $N = 2$, we recover 
Equation \ref{eq3}.  

Intuitively, by prescribing several scaling exponents, $\zeta_q$ (which are known apriori from observational data), 
it is possible to solve for $d_n$ from the overdetermined system of equations (Equation \ref{eq6}). These solved parameters, $d_
n$, along with other easily derivable (from the given anchor points and $d_n$) parameters ($a_n,  c_n,  e_n$ and $f_n$) 
in turn can be used to construct multiaffine signals. For example, solving for the values quoted by Frisch \cite{Frisch}:
$\zeta_2 = 0.70, \zeta_3 = 1, \zeta_4 = 1.28, \zeta_5 = 1.53, \zeta_6 = 1.77, \zeta_7 = 2.01$ and $\zeta_8 = 2.23$, 
along with $\zeta_1 = 0.33$ (corresponding to $D = 5/3$), yields the stretching factors $\left| d_n \right| = 
0.887, 0.676$. There are altogether eight possible sign combinations for the above stretching parameter magnitudes 
and all of them can potentially produce multiaffine fields with the aforementioned scaling exponents. However, all of 
them might not be the ``right" candidate from the LES-performance perspective. Rigorous a-priori and a-posteriori testing of 
these Multiaffine SGS models is needed to elucidate this issue (see Section \ref{Sec5}).

We repeated our previous numerical experiment with the stretching parameters $d_1 = -0.887$ and $d_2 = 0.676$. Figure 
3a shows the measured values (ensemble averaged over one hundred realizations) of the scaling exponents $\zeta_q$ upto 
$12^{th}$ order. For comparison we have also shown the theoretical values computed directly from Equation \ref{eq6} 
(dashed line). A model proposed by She and L\'{e}v\^{e}que \cite{She} based on a hierarchy of fluctuation structures 
associated with the vortex filaments is also shown for comparison (dotted line). We chose this particular model because of its 
remarkable agreement with experimental data. The She and L\'{e}v\^{e}que model predicts: $\zeta_q = \frac{q}{9} + 2 - 2(\frac{2}{3})^
{ q \mathord{\left/ {\vphantom {1 3}} \right. \kern-\nulldelimiterspace} 3}$. Figure 3b shows the pdfs of the increments,
which are quite similar to what is observed in real turbulence -- for large $r$ the pdf is near Gaussian while for 
smaller $r$ it becomes more and more peaked at the core with high tails (see also Figure 7b for the variation of flatness 
factors of the pdfs of increments with distance $r$). 

\medskip
\section{\label{Sec4} Fractal Calculus and Subgrid-Scale Modeling} 

In the case of an incompressible fluid, the spatially filtered Navier-Stokes equations are:
\begin{subequations}
\begin{equation}\label{NS1}
\frac{\partial \tilde {u}_m }{\partial x_m } = 0,
\end{equation}
\begin{eqnarray}\label{NS2}
\frac{\partial \tilde {u}_m }{\partial t} + \tilde {u}_n \frac{\partial 
\tilde {u}_m }{\partial x_n } = - \frac{\partial }{\partial x_n }\left[ 
{\frac{\tilde {p}}{\rho }\delta _{mn} + \tau _{mn} } \right] + \nu \nabla^2 \tilde {u}_m \\ 
\mbox{} m,n = 1,2,3. \nonumber
\end{eqnarray}
\end{subequations}
\noindent
where $t$ is time, $x_n$ is the spatial coordinate in the $n$-direction, $u_n$ is the velocity component in the 
$n$-direction, $p$ is the dynamic pressure, $\rho$ is the density and $\nu$ is the molecular viscosity of the fluid. 
The tilde $\tilde{(\hspace{0.1in})}$ denotes the filtering operation, using a filter of characteristic width $\Delta$ 
\cite{Footnote1}. 
These filtered equations are now amenable to numerical solution (LES) on a grid with mesh-size of order $\Delta$,
considerably larger than the smallest scale of motion (the Kolmogorov scale). However, the SGS stress tensor $\tau_{mn}$ in 
Equation \ref{NS2}, defined as
\begin{equation}\label{}
\tau _{mn} = \widetilde{u_m u_n } - \tilde {u}_m \tilde {u}_n \\ 
\end{equation}
is not known. It essentially represents the contribution of unresolved scales (smaller than $\Delta$) to the total momentum
transport and must be parameterized (via a SGS model) as a function of the resolved velocity field. Due to strong 
influence of the SGS parameterizations on the dynamics of the resolved turbulence, considerable research efforts 
have been made during the past decades and several SGS models have been proposed (see \cite{Meneveau-Katz,Sagaut} for reviews). 
The Eddy-viscosity model (\cite{Smagorinsky}) and its variants (e.g., the Dynamic model \cite{Germano}, the 
Scale-Dependent Dynamic model \cite{Porte1}) are perhaps the most widely used SGS models. They parameterize the 
SGS stresses as being proportional to the resolved velocity gradients. These SGS models and other standard models 
(e.g., Similarity, Nonlinear, Mixed models) postulate the form of the SGS stress tensors rather than the structure of 
the SGS fields (\cite{Mazzino}). Philosophically a very different approach would be to explicitly reconstruct the 
subgrid-scales from given resolved scale values (by exploiting the statistical structures of the unresolved turbulent 
fields) using a specific mathematical tool (e.g., the Fractal Interpolation Technique) and  subsequently estimate 
the relevant SGS tensors necessary for LES. The Fractal model of \cite{Scotti97,Scotti99} and our proposed Multiaffine 
model basically represent this new class of SGS modeling, also known as the ``direct modeling of SGS turbulence'' 
(\cite{Meneveau-Katz,Sagaut,Doma}).

In Section \ref{Sec3}, we have demonstrated that FIT could be effectively used to generate synthetic turbulence 
fields with desirable statistical properties. In addition, Barnsley's rigorous fractal calculus offers the ability to analytically 
evaluate any statistical moment of these synthetically generated fields, which in turn could be used for SGS modeling.  
Detailed discussion of the fractal calculus is beyond the scope of this paper. Below, we briefly summarize the 
equations most relevant to the present work. Let us first consider the moment integral:
$U_{l,m} = \int\limits_{0}^{1} {x^m\left[ {u\left( x \right)} \right]^ldx}.$ 
In the present context $(x_0 = 0, x_1 = 0.5 \mbox{ and } x_2 = 1)$, this moment integral could be viewed as $2\Delta$
filtering $(\Delta = 0.5)$ with the Top Hat filter 
$\mbox{[ i.e., } F_\Delta \left( {x} \right) = 1/\Delta,\mbox{ if }\left| {x} \right| < \Delta/2 \mbox{ and } F_\Delta(x) = 0, \mbox{ otherwise } \mbox{]}$.
For instance, the 1-D component of the SGS stress tensor reads as: 
\begin{subequations}\begin{eqnarray}
\tau = & \widetilde{uu} & - \tilde {u} \tilde {u}\\
     = & U_{2,0}        & - U_{1,0}U_{1,0}
\end{eqnarray}\end{subequations}

Barnsley (\cite{Barn86}) proved that for the fractal interpolation IFS (Equation \ref{eq1}), the moment integral becomes: 
\begin{widetext}
\begin{subequations}
\begin{equation}\label{}
U_{l,m} = \frac{\left( {\sum\limits_{j = 0}^{m - 1} {U_{l,j} } \left( 
{{\begin{array}{*{20}c}
 m \hfill \\
 j \hfill \\
\end{array} }} \right)\sum\limits_{n = 1}^2 {a_n^{j + 1} d_n e_n^{m - j} } + 
\sum\limits_{p = 0}^{l - 1} {\sum\limits_{j = 0}^{l + m - p} {K\left( 
{l,m,p,j} \right)U_{p,j} } } } \right)}{\left( {1 - \sum\limits_{n = 1}^2 
{a_n^{m + 1} d_n^l } } \right)} \\
\end{equation}
\noindent
where,
\begin{equation}
\sum\limits_{n = 1}^2 {\left( {{\begin{array}{*{20}c}
 l \hfill \\
 p \hfill \\
\end{array} }} \right)a_n \left( {c_n x + f_n } \right)^{l - p}d_n^p \left( 
{a_n x + e_n } \right)^m = \sum\limits_{j = 0}^{l + m - p} {K\left( 
{l,m,p,j} \right)x^j} } 
\end{equation}
\end{subequations}
\end{widetext}
After some algebraic manipulations, the SGS stress equation at node $x_i$ becomes: 
\begin{eqnarray}\label{eq11}
\tau_{i} = & \alpha_{0}\tilde {u}_{i-1}^2 + \alpha_{1}\tilde {u}_i^2 + \alpha_{2}\tilde {u}_{i+1}^2 + \nonumber \\ 
		   & \alpha_{3}\tilde {u}_{i-1} \tilde {u}_i + \alpha_{4}\tilde {u}_i \tilde {u}_{i+1} + \alpha_{5} \tilde {u}_{i+1} \tilde {u}_{i-1}  
\end{eqnarray}
We would like to point out that the coefficients $\alpha_k$ are sole functions of the stretching factors $d_n$. 
In other words, if one can specify the values of $d_n$ in advance, the SGS stress ($\tau$) could be explicitly written
in terms of the coarse-grained (resolved) velocity field ($\tilde {u}_i$) weighted according to weights $\alpha_k$ 
uniquely determined by $d_n$. 
In Table~\ref{tab:T1}, we have listed the $\alpha_k$ values corresponding to eight stretching factor combinations, 
$\left| d_n \right| = 0.887, 0.676$. 
It is evident that any two combinations ($d_1,d_2$) and ($d_2,d_1$) are simply ``mirror'' images of each 
other in terms of $\alpha_k$. Thus, only four distinct Multiaffine SGS models (M1, M2, M3 and M4) could be formed from 
the aforementioned eight $\left| d_n \right|$ combinations and in each case the orderings could be chosen at random with 
equal probabilities. 
In this table, we have also included the Fractal model of \cite{Scotti97,Scotti99} and the Similarity model of 
\cite{Bard} in expanded form similar to the Multiaffine models (see the Appendix for more information on standard SGS 
models). The Multiaffine models and the Fractal model differ slightly in terms of filtering  operation. 
\cite{Scotti97,Scotti99} performed filtering at a scale $\Delta$ (see Equation \ref{Scotti}), whereas in the case of Similarity 
model, \cite{Liu} found that it is more appropriate to filter at $2\Delta$. For the Multiaffine models, we also chose 
to employ $2\Delta$ filtering scale. 

One noticable feature in Table \ref{tab:T1} is that some combinations of $d_n$ result in strongly asymmetric 
weights $\alpha_k$. As an example, 
in the case of M4 with $d_1 = +0.676$ and $d_2 = +0.887$, $|\alpha_0| \gg |\alpha_2|$ and $|\alpha_3| \gg |\alpha_4|$. 
This means that the SGS stress at any node $x_i$ would have more weight from the resolved velocity at node $x_{i-1}$ 
than node $x_{i+1}$. One would expect that such an asymmetry could have serious implication in terms of SGS model 
performance. 

In the following section, we will attempt to address this issue among others by evaluating several SGS 
models via the a-priori analysis approach. 

\begin{widetext}
\begin{table*}
\caption{\label{tab:T1}The Multiaffine, Fractal and Similarity SGS models in expanded form and their corresponding 
coefficients for the computation of SGS stresses according to Equation \ref{eq11}.}
\begin{ruledtabular}
\begin{tabular}{lcccccccc}
Model & $(d_1,d_2)$	& Filter Width & $\alpha_0$ & $\alpha_1$ & $\alpha_2$ & $\alpha_3$ & $\alpha_4$ & $\alpha_5$ \\
\hline
Multiaffine (M1) & $(-0.887,+0.676)$ & $2\Delta$ & $0.218$ & $0.204$ & $0.050$ & $-0.372$ & $-0.036$ & $-0.065$ \\
                 & $(+0.676,-0.887)$ & $2\Delta$ & $0.050$ & $0.204$ & $0.218$ & $-0.036$ & $-0.372$ & $-0.065$ \\
Multiaffine (M2) & $(+0.887,-0.676)$ & $2\Delta$ & $0.030$ & $0.248$ & $0.261$ & $-0.018$ & $-0.479$ & $-0.043$ \\
                 & $(-0.676,+0.887)$ & $2\Delta$ & $0.261$ & $0.248$ & $0.030$ & $-0.479$ & $-0.018$ & $-0.043$ \\
Multiaffine (M3) & $(-0.887,-0.676)$ & $2\Delta$ & $0.144$ & $0.220$ & $0.133$ & $-0.230$ & $-0.209$ & $-0.057$ \\
                 & $(-0.676,-0.887)$ & $2\Delta$ & $0.133$ & $0.220$ & $0.144$ & $-0.209$ & $-0.230$ & $-0.057$ \\
Multiaffine (M4) & $(+0.887,+0.676)$ & $2\Delta$ & $0.064$ & $0.319$ & $0.262$ & $-0.121$ & $-0.517$ & $-0.007$ \\
                 & $(+0.676,+0.887)$ & $2\Delta$ & $0.262$ & $0.319$ & $0.064$ & $-0.517$ & $-0.121$ & $-0.007$ \\
Fractal          & $(-0.794,+0.794)$ & $\Delta$  & $0.127$ & $0.221$ & $0.026$ & $-0.322$ & $-0.120$ & $+0.069$ \\
                 & $(+0.794,-0.794)$ & $\Delta$  & $0.026$ & $0.221$ & $0.127$ & $-0.120$ & $-0.322$ & $+0.069$ \\
Similarity       & NA                & $2\Delta$ & $0.188$ & $0.250$ & $0.188$ & $-0.250$ & $-0.250$ & $-0.125$ \\
\end{tabular}
\end{ruledtabular}
\end{table*}
\end{widetext}

\section{\label{Sec5} Evaluation of SGS Models: A-priori Analysis Approach} 

The SGS models and their underlying hypotheses can be evaluated by two approaches: a-priori testing and a-posteriori 
testing (terms coined by \cite{Piomelli}). In a-posteriori testing, LES computations are actually performed with 
proposed SGS models and validated against reference solutions (in terms of mean velocity, scalar and stress 
distributions, spectra etc.). However, owing to the multitude of factors involved in any numerical simulation (e.g.,
numerical discretizations, time integrations, averaging and filtering), a-posteriori tests in general do not provide 
much insight about the detailed physics of the newly implemented SGS models \cite{Meneveau-Katz,Sagaut}. A 
complementary  and perhaps more fundamental approach \cite{Meneveau-Katz} would be to use high-resolution model (Direct 
Numerical Simulation or DNS), experimental or field observational data to compute the ``real'' and modeled SGS tensors 
directly from their definitions and compare them subsequently. This approach, widely known as the a-priori analysis, 
does not require any actual LES modeling and is theoretically more tractable. 
In this work we focused on comparing the performance of SGS models via the a-priori analysis. We
strictly followed the 1-D apriori analysis approach of \cite{Meneveau,Oneil,Porte2}. To highlight the caveats of the proposed 
and several existing SGS models, we performed an extensive inter-model comparison study. This exercise also helped
selecting the ``right'' combination of stretching factors for the Multiaffine SGS models.

In general, the correlation between real ($\tau^{Real}$) and modeled ($\tau^{Model}$) SGS stresses is considered to 
be a good indicator of the expected performance of a proposed SGS model. Another crucial indicator is the 
so called SGS energy dissipation rate ($\Pi$): 
\begin{eqnarray}
\Pi &= &-\tau_{ij}\tilde{S}_{ij} \\
         &\approx &-\frac{15}{2}\tau\frac{\partial\tilde{u}}{\partial x} \nonumber \mbox{~(1-D approximation)}
\end{eqnarray} 
In the inertial range, the SGS energy dissipation rate is the most influential factor affecting the dynamical 
evolution of the resolved kinetic energy \cite{Meneveau-Katz}. On average, $\Pi$ is positive, representing a net
drain of resolved kinetic energy into unresolved motion. Intermittent negative values of $\Pi$, known as
``backscatter'', implies energy transfer from SGS to resolved scales. 
Unfortunately, a high correlation between real and modeled SGS stress (or SGS energy dissipation rate) is not a sufficient 
condition for the success of a proposed LES SGS model, although it is a highly desirable feature \cite{Liu,Meneveau,Sagaut}.
  
We primarily made use of an extensive atmospheric boundary layer (ABL) turbulence dataset (comprising of 
fast-response sonic anemometer data) collected by various researchers from the Johns Hopkins University, the University 
of California-Davis and the University of Iowa during Davis 1994, 1995, 1996, 1999 and Iowa 1998 field studies. 
Comprehensive description of these field experiments (e.g., surface cover, fetch, instrumentation, sampling frequency) 
can be found in \cite{Pahlow}. 
We further augmented this dataset with nocturnal ABL turbulence data from CASES-99 (Cooperative Atmosphere-Surface 
Exchange Study 1999), a cooperative field campaign conducted near Leon, Kansas during October 1999 \cite{Poulos}. For 
our analyses four levels (1.5, 5, 10 and 20m) of sonic anemometer data from the 60m tower and the adjacent mini-tower 
collected during two intensive observational periods (nights of October $17^{th}$ and $19^{th}$) were considered [the 
sonic anemometer at 1.5m was moved to 0.5m level on October $19^{th}$]. Briefly, the collective attributes of the field 
dataset explored in this study are as follows: (i) surface cover: bare soil, grass and beans; (ii) sampling frequency: 
18 to 60 Hz; (iii) sampling period: 20 to 30 minutes; (iv) sensor height ($z$): 0.5 to 20m; and (v) atmospheric 
stability ($z/L$, $L$ is the local Obukhov length): ~0 (neutral) to ~10 (very stable). 

ABL field measurements are seldom free from mesoscale disturbances, wave activities, nonstationarities etc. The 
situation could be further aggravated by several kinds of sensor errors (e.g., random spikes, amplitude resolution error, 
drop outs, discontinuities etc.). Thus, stringent quality control and preprocessing of field data is of utmost 
importance for any rigorous statistical analysis. Our quality control and preprocessing strategies are qualitatively 
similar to the suggestions of \cite{Mahrt1}. Specifically, we follow these steps:

(1) Visual inspection of individual data series for detection of spikes, amplitude resolution error, drop 	      
outs and discontinuities. Discard suspected data series from further analyses.
                                     	
(2) Adjust for changes in wind direction by aligning sonic anemometer data using 60 seconds local 
averages of the longitudinal and transverse component of velocity.                               	

(3) Partitioning of turbulent-mesoscale motion using discrete wavelet transform (Symmlet-8 wavelet) with a gap scale
\cite{Mahrt2} of 100 seconds.

(4) Finally, to check for nonstationarities of the partitioned series, we performed the following step: we subdivided 
each series in 6 equal intervals and computed the standard deviation of each sub-series ($\sigma_i,~i=1:6$). If 
$max(\sigma_i) / min(\sigma_i) > 2$, the series was discarded.                                      

After all these quality control and preprocessing steps, we were left with only 358 ``reliable'' series for a-priori 
analyses. These streamwise velocity series were filtered with a Top-Hat filter ($\Delta =$ 1,2,4 or 8m) and downsampled
on the scale of the LES grid ($\Delta$) to obtain the resolved velocity field $\tilde{u}_i$ \cite{Footnote2}.  
In a similar way, the streamwise SGS stress, $\tau^{Real}$, was computed from its definition (Equation 9a). Filtering 
operations were always performed in time and interpreted as 1-D spatial filtering in the streamwise direction by means of
Taylor's frozen flow hypothesis. The spatial derivatives were also computed from the time derivatives by invoking 
Taylor's hypothesis: $\frac{\partial}{\partial x} = -{\frac{1}{<u>}}{\frac{\partial}{\partial t}}$, where, $<u>$ is the 
mean streamwise velocity. 

In Figure 4 representative realizations of the real and several modeled SGS stresses are presented. The modeled SGS 
stresses, $\tau^{Model}$ were computed from the definitions given in the Appendix.
Along the same lines, the real and modeled SGS energy dissipation rates (Figure 5) were calculated according to   
$\Pi^{Real}  = -\frac{15}{2}\tau^{Real}\frac{\partial \tilde {u}}{\partial x}$ and 
$\Pi^{Model} = -\frac{15}{2}\tau^{Model}\frac{\partial \tilde {u}}{\partial x}$, respectively.
The SGS model constants like $C_S$ of the Smagorinsky model or $C_L$ of the Similarity model (see Appendix) were obtained by matching
the mean real and modeled SGS energy dissipation rates \cite{Meneveau,Oneil,Porte2}. For consistency, the same 
procedure was followed for the SGS-kinetic-energy-based model, Fractal model and Multiaffine models. In other words, we always ensured that 
$<-\frac{15}{2}\tau^{Real}\frac{\partial \tilde {u}}{\partial x}>~=~<-\frac{15}{2}\tau^{Model}\frac{\partial \tilde {u}}{\partial x}>$.
Note that, this procedure has no effect on the correlation results presented below. In actual simulations these model
coefficients could be obtained dynamically following the approach of \cite{Germano,Lilly}. 

From Figures 4 and 5 it is visually evident that both Similarity and Multiaffine models capture the variability of 
the SGS stress and energy dissipation rates reasonably well. On the other hand, the performances of the Smagorinsky and 
SGS-kinetic-energy-based models are very poor. Note that, the Smagorinsky model assumes that the trace of the SGS tensor is subtracted 
from the tensor, which is not feasible in 1-D a-priori analysis \cite{Meneveau}. Thus, direct magnitude-wise comparison
between real and the Smagorinsky model based SGS stress or dissipation energy is not possible. However, this does not 
prevent us from quantifying the performance by correlation coefficient. Moreover, the Smagorinsky model is by 
construction fully dissipative. Hence, this model is unable to reproduce the backscatter effects (see Figure 5b), 
which do occur in the real SGS dissipation series (Figure 5a).

In Table \ref{tab:T2}, for $\Delta =$ 1m, we show the correlation between the real and modeled SGS stress and energy dissipation rates. 
The standard deviations are given in parentheses. The model M3 is significantly better than any other Multiaffine model 
and this could only be attributed to its near-symmetric stencil structure (see Table 1). This resolves our previous 
dilemma regarding the selection of one Multiaffine SGS model from a class of four. From here on, we'll only report 
results for M3 and will identify it as the Multiaffine model. 
\begin{table}
\caption{\label{tab:T2} Average correlation between observed and modeled SGS stresses and energy dissipation rates 
$(\Delta = 1$m). The results are based on 358 ABL turbulent velocity series measured during several field campaigns. 
The quantities in the parenthesis represent standard deviation.}
\begin{ruledtabular}
\begin{tabular}{lcc}
                 & Corr$(\tau^{Real},\tau^{Model})$& Corr$(\Pi^{Real},\Pi^{Model})$ \\
\hline
Smagorinsky 	 & $0.25(0.09)$     & $0.41(0.17)$\\
Similarity  	 & $0.49(0.10)$     & $0.76(0.15)$\\
SGS-KE	  		 & $0.23(0.08)$     & $0.42(0.17)$\\
Fractal			 & $0.33(0.05)$     & $0.61(0.06)$\\
Multiaffine (M1) & $0.44(0.05)$     & $0.71(0.05)$\\
Multiaffine (M2) & $0.40(0.05)$     & $0.68(0.06)$\\
Multiaffine (M3) & $0.49(0.05)$     & $0.77(0.05)$\\
Multiaffine (M4) & $0.42(0.05)$     & $0.70(0.05)$\\
\end{tabular}
\end{ruledtabular}
\end{table}

Next, in Figure 6, we plot the mean correlation between real and modeled SGS stress and energy dissipation rates for $\Delta$
= 1,2,4 and 8m. As anticipated, for all the models, the correlation decreases with increasing filtering scale. Also, the correlation
of real vs. model SGS energy dissipation rates is usually higher compared to the SGS stress scenario, as noticed by other 
researchers.   

It is expected that in the ABL the scaling exponent values ($\zeta_q$) would deviate from the values reported in 
\cite{Frisch} due to near-wall effect. 
This means that the stretching factors $d_n$ based on the $\zeta_q$ values we used in this work, are possibly in 
error. Nevertheless, the overall performance of the Multiaffine model is beyond our expectations. It remains to be seen
how the proposed SGS scheme will perform in a-posteriori analysis and such work is currently in progress.

\section{Passive Scalar}

Our scheme could be easily extended to synthetic passive-scalar (any diffusive component in a fluid flow that has no 
dynamical effect on the fluid motion itself, e.g., a pollutant in air, temperature in a weakly heated flow, a dye mixed 
in a turbulent jet or moisture mixing in air \cite{War,Sigg}) field generation. The statistical and dynamical 
characteristics (anisotropy, intermittency, pdfs etc.) of passive-scalars are surprisingly different from the 
underlying turbulent velocity field \cite{War,Sigg}. For example, it is even possible for the passive-scalar field to 
exhibit intermittency in a purely Gaussian velocity field \cite{War,Sigg}. Similar to the K41, neglecting intermittency,
the Kolmogorov-Obukhov-Corrsin (KOC) hypothesis predicts that at high Reynolds and Peclet numbers, the $q^{th}$-order 
passive-scalar structure function will behave as: $\left\langle {\left| {\theta\left( {x + r} \right) - \theta\left( x 
\right)} \right|^q} \right\rangle \sim r^{\frac{q}{3} }$ in the inertial range. Experimental observations reveal that 
analogous to turbulent velocity, passive-scalars also exhibit anomalous scaling (departure from the KOC scaling). 
Observational data also suggest that passive-scalar fields are much more intermittent than velocity fields and result 
in stronger anomaly \cite{War,Sigg}. 

To generate synthetic passive-scalar fields, we need to determine the stretching parameters $d_1$ and $d_2$ from 
prescribed scaling exponents, $\zeta_q$. Unlike the velocity scaling exponents, the published values (based on 
experimental observations) of higher-order passive-scalar scaling exponents display significant scatter. Thus for our 
purpose, we used the predictions of a newly proposed passive-scalar model \cite{Feng}: $\zeta _q = 2 + \left( {\frac{8}
{9}} \right)^2 - 2\left( {\frac{3}{4}} \right)^{q \mathord{\left/ {\vphantom {q 6}} \right. \kern-\nulldelimiterspace} 
6} - \left( {\frac{8}{9}} \right)^2\left( {\frac{7}{16}} \right)^{q \mathord{\left/ {\vphantom {q 2}} \right. 
\kern-\nulldelimiterspace} 2}$. This model based on the hierarchical structure theory of \cite{She} shows reasonable 
agreement with the observed data. Moreover, unlike other models, this model manages to predict that the scaling 
exponent $\zeta_q$ is a nondecreasing function of $q$. Theoretically, this is crucial because, otherwise, if $
\zeta_q \to -\infty$ as $q \to +\infty$, the passive-scalar field cannot be bounded \cite{Frisch,Feng}. 

Employing Equation (\ref{eq6}) and the scaling exponents (upto $8^{th}$-order) predicted by the above model, we get the 
following stretching factors: $\left| d_n \right| = 0.964, 0.606$. We again repeated the numerical experiment of section 
\ref{Sec3} and selected the stretching parameter combination: $d_1 = -0.964$ and $d_2 = 0.606$. Like before, we compared 
the estimated [using Equation (4)] scaling exponents from one hundred realizations 
with the theoretical values [from Equation (6)] and the agreement was found to be highly satisfactory. To check whether 
a generated passive-scalar field ($d_1 = -0.964$, $d_2 = 0.606$) possesses more non-Gaussian characteristics than its 
velocity counterpart ($d_1 = -0.887$, $d_2 = 0.676$), we performed a simple numerical experiment. We generated both 
the velocity and passive-scalar fields from identical anchor points and also computed the corresponding flatness factors, 
$K$, as a function of distance $r$ (see Figure 7b). 
Comparing Fig 7a with Fig 3b and also from Fig 7b, one could conclude that the passive-scalar field exhibits stronger 
non-Gaussian behavior than the velocity field, in accord with the literature. 

\section{Concluding Remarks}

In this paper, we propose a simple yet efficient scheme to generate synthetic turbulent velocity and passive-scalar 
fields. This method is competitive with most of the other synthetic turbulence emulator schemes (e.g., \cite{Vicsek,
Benzi,Juneja,Biferale,Bohr}) in terms of capturing small-scale properties of turbulence and scalars (e.g., multiaffinity 
and non-Gaussian characteristics of the pdf of velocity and scalar increments). Moreover, extensive a-priori analyses of 
field measurements unveil the fact that this scheme could be effectively used as a SGS model in LES.
Potentially, the proposed Multiaffine SGS model can address two of the unresolved issues in LES: it 
can systematically account for the near-wall and atmospheric stability effects on the SGS dynamics. 
Of course, this would require some kind of universal dependence of the scaling exponents on both wall-normal distance 
and stability. Quest for this kind of universality has began only recently \cite{Ruiz,Aivalis}.

\begin{acknowledgments}
We thank Alberto Scotti, Charles Meneveau, Andreas Mazzino, Venugopal Vuruputur and Boyko Dodov for useful discussions. 
The first author is indebted to Jacques L\'{e}vy-V\'{e}hel for his generous help. This work was partially funded by 
NSF and NASA grants. One of us (SB) was partially supported by the Doctoral Dissertation Fellowship from the University 
of Minnesota. All the computational resources were kindly provided by the Minnesota Supercomputing Institute. All 
these supports are greatly appreciated.
\end{acknowledgments}

\appendix*
\section{}

\begin{subequations}
The standard Smagorinsky eddy-viscosity model is of the form:
\begin{equation}
\tau^{Smag}_{ij} = -2(C_S\Delta)^2|\tilde {S}|\tilde {S}_{ij}
\end{equation}
where,
\begin{eqnarray*}
\tilde {S}_{ij} = \frac{1}{2}( {\frac{\partial \tilde {u}_i }{\partial x_j } + 
\frac{\partial \tilde {u}_j }{\partial x_i }}) 
\end{eqnarray*}
is the resolved strain rate tensor and
\begin{eqnarray*}
|\tilde {S}| = ({2\tilde {S}_{ij} \tilde {S}_{ij} })^{1/2}
\end{eqnarray*}
is the magnitude of the resolved strain rate tensor. $C_S$ is the so-called Smagorinsky coefficient. 

For 1-D surrogate SGS stress: 
\begin{eqnarray*}
\tilde {S}_{11} = \frac{\partial \tilde {u}}{\partial x} 
\end{eqnarray*}
Further, by assuming that the smallest scales of the resolved motion are isotropic, the following equality holds 
\cite{Monin}:
\begin{eqnarray*}
<\tilde {S}_{ij} \tilde {S}_{ij}> = \frac{15}{2} <\tilde {S}_{11}^2>
\end{eqnarray*}
Employing this assumption for the instantaneous fields, we can write
\begin{eqnarray*}
|\tilde {S}| = ({2\tilde {S}_{ij} \tilde {S}_{ij} })^{1/2} \approx \sqrt {15}| {\frac{\partial \tilde {u}}{\partial x}}| 
\end{eqnarray*}
Hence, the Smagorinsky SGS stress equation becomes:
\begin{equation}
\tau ^{Smag} = - 2({C_S \Delta })^2\sqrt {15} |{\frac{\partial \tilde {u}}{\partial x}}|
({\frac{\partial \tilde {u}}{\partial x}}) 
\end{equation} 
\end{subequations}
The second model we considered is the Similarity Model \cite{Bard,Liu}. 
\begin{subequations}
\begin{equation}
\tau^{Siml}_{ij} = C_L(\overline {\tilde {u}_i \tilde {u}_j} - {\overline {\tilde{u}}}_i~{\overline {\tilde{u}}}_j)\\
\end{equation}
The overbar denotes explicit filtering with a filter of width $\gamma\Delta$ (usually $\gamma = 2$). $C_L$ is the 
Similarity model coefficient.

The 1-D surrogate SGS stress could be simply written as:
\begin{equation}
\tau^{Siml} = C_L(\overline {\tilde {u} \tilde {u}} - {\overline {\tilde{u}}}~{\overline {\tilde{u}}})\\
\end{equation}
Now, for 2$\Delta$ filtering this equation becomes:
\begin{equation}
\tau^{Siml} = C_L[({\frac{\tilde {u}_{i-1}^2 + 2\tilde {u}_i^2 + \tilde {u}_{i+1}^2 }{4}}) \\ 
                   - ({\frac{\tilde {u}_{i-1} + 2\tilde {u}_i + \tilde {u}_{i+1} }{4}})^2] 
\end{equation}
which on further simplification leads to the expression in Table \ref{tab:T1}:  
\begin{eqnarray*}
\tau^{Siml} = C_L[0.188\tilde{u}_{i-1}^2 + 0.25\tilde{u}_i^2 + 0.188\tilde{u}_{i+1}^2 \\
            - 0.25\tilde{u}_{i-1}\tilde{u}_i - 0.25\tilde{u}_i\tilde{u}_{i+1} - 0.125\tilde{u}_{i+1}\tilde{u}_{i-1}]
\end{eqnarray*}
\end{subequations}
Next, we consider a SGS model based on the SGS kinetic energy $(q^2)$ \cite{Meneveau,Wong}:
\begin{subequations}
\begin{eqnarray}
\tau^{Stke} = -2C_K^2\Delta\sqrt {|\tilde {q^2}|}{\frac{\partial \tilde {u}}{\partial x}}\\
\mbox{where,~} q^2 = 3(u-\tilde{u})^2 \nonumber
\end{eqnarray}
\end{subequations}
Here, $C_K$ is the SGS model coefficient.

In the case of the Fractal model of \cite{Scotti97,Scotti99}, the unknown subgrid stress ($\tau$) produced by  
synthetic fractal field around any grid point $x_i$ can be written as:
\begin{subequations}
\begin{eqnarray}\label{Scotti} 
\tau_i^{Frac} & = & \int \limits_{\frac{1}{4}}^{\frac{3}{4}} [u(x)]^2dx - [\int \limits_{\frac{1}{4}}^{\frac{3}{4}} u(x)dx]^2 \\
              & = & \frac{1}{12}(\delta_i\tilde{u})^2 + \frac {d_i(8-3d_i^2)}{48}\delta_i^2\tilde{u}\delta_i\tilde{u} + \nonumber \\
              &   & \frac{1+15d_i^2-24d_i^4+12d_i^6}{192(1-d_i^2)}(\delta_i^2\tilde{u})^2.
\end{eqnarray}
\end{subequations}
where, $\delta_i\tilde{u} = (\tilde{u}_{i+1} - \tilde{u}_{i-1})/2,~\delta_i^2\tilde{u} = \tilde{u}_{i+1} - 2\tilde{u}_i+
\tilde{u}_{i-1},$ and $d_i = \pm2^{-1/3}$.

\begin{figure}[ht]
\includegraphics[width=15pc]{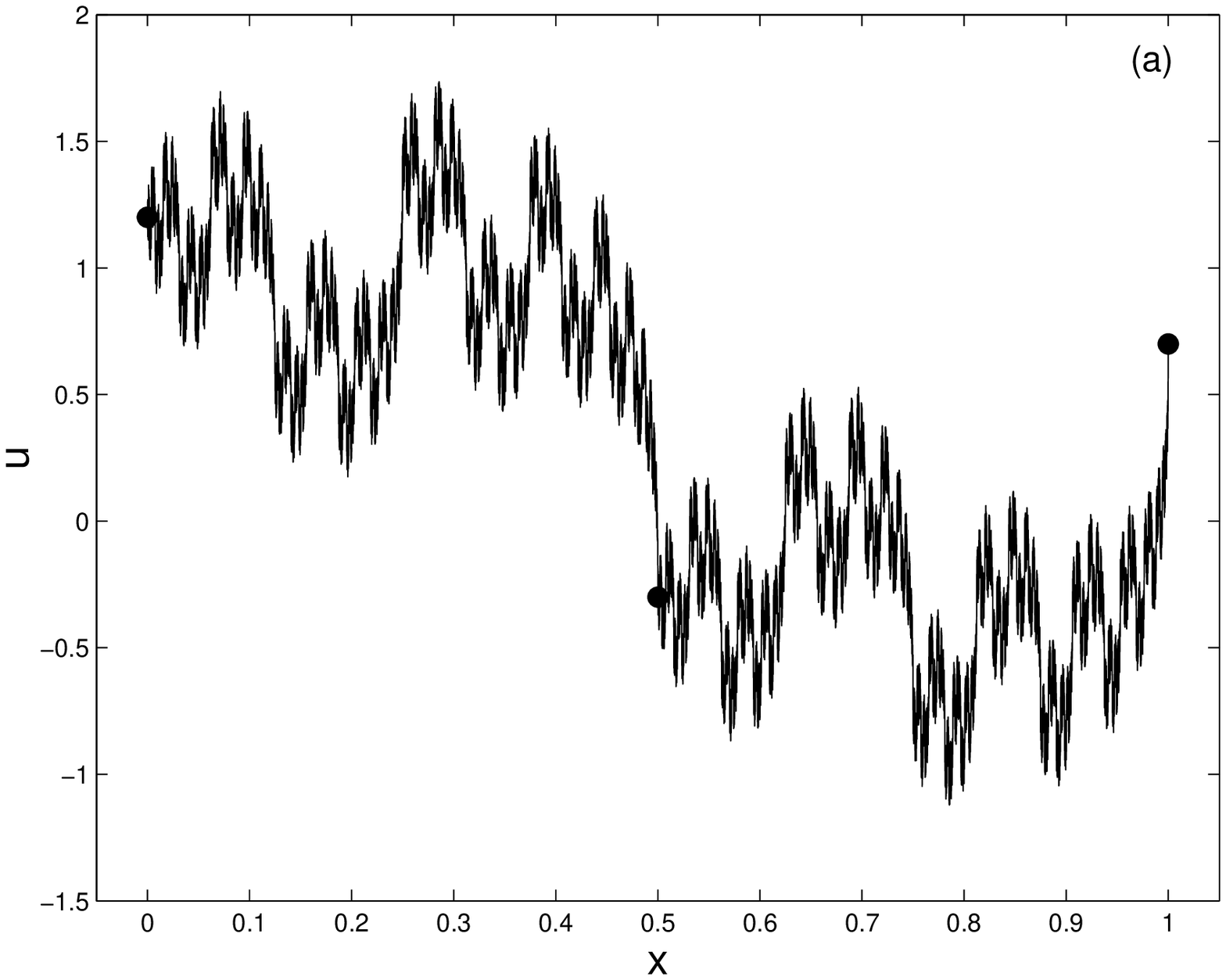}
\includegraphics[width=15pc]{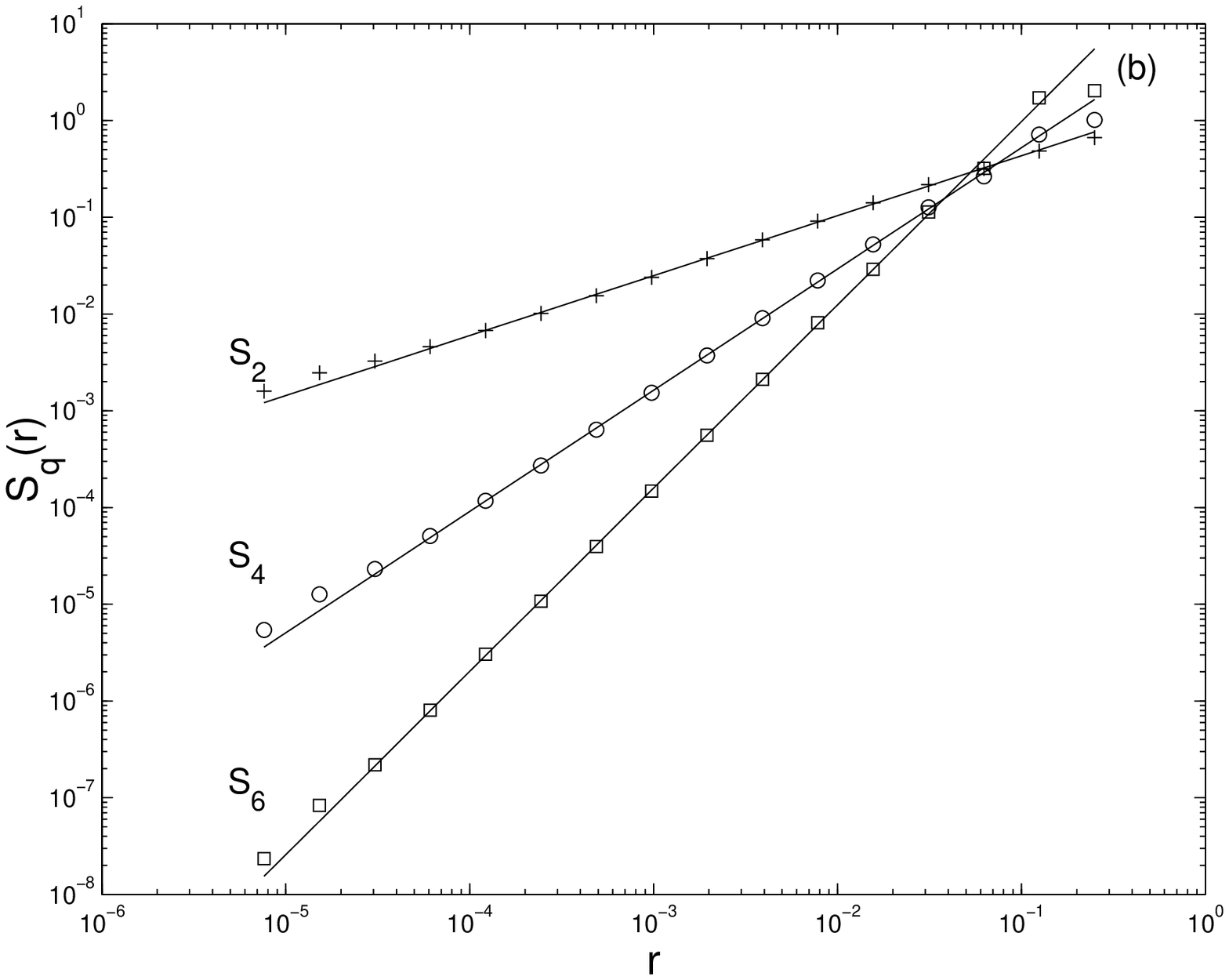}
\includegraphics[width=15pc]{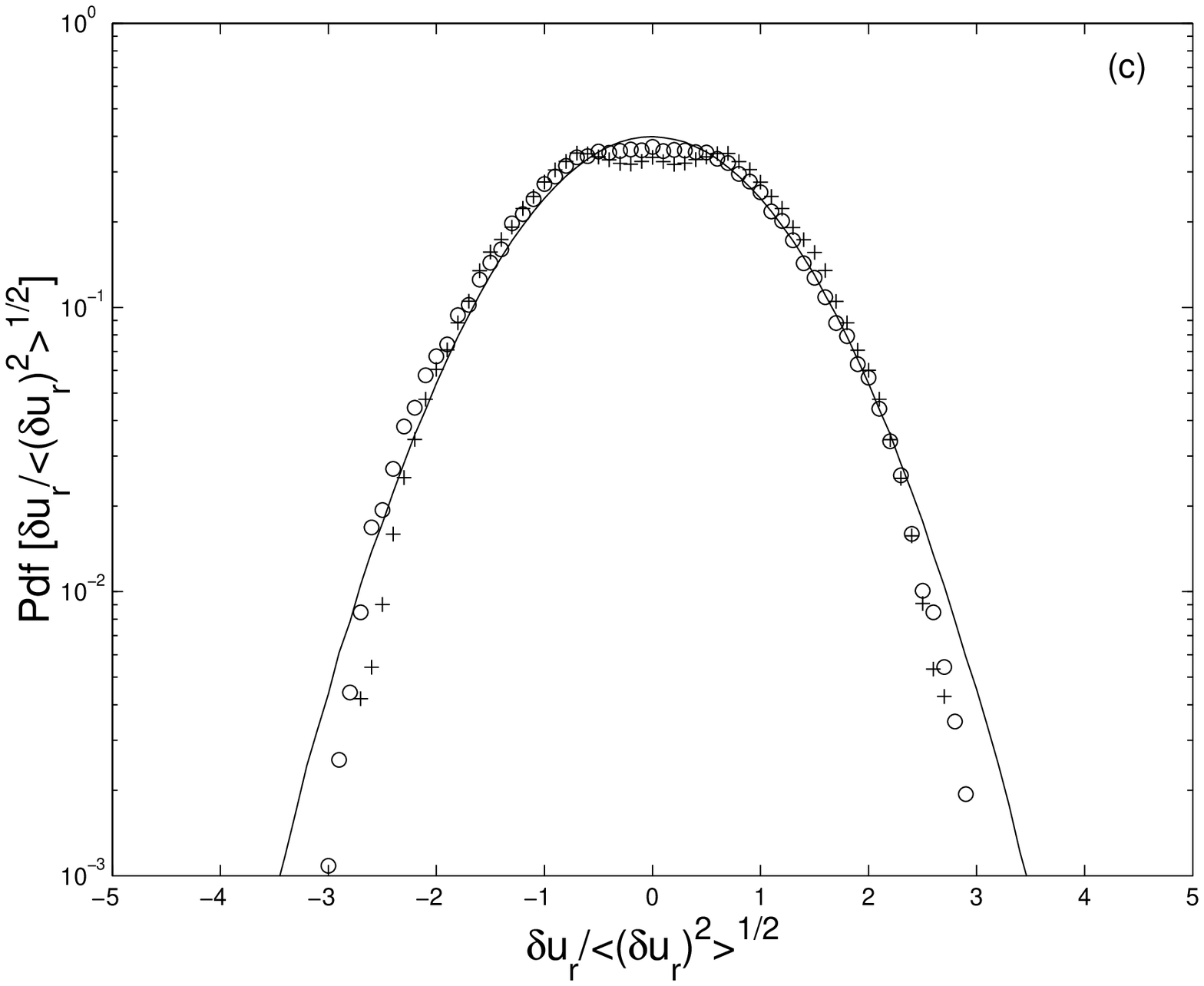}
\caption{(a) A synthetic turbulence series of fractal dimension $D = 5/3$. The black dots denote initial 
interpolating points. (b) Structure functions of order 2, 4 and 6 (as labeled) computed from the series in Figure 1a. 
The slopes $(\zeta_q)$ corresponding to this particular realization are 0.62, 1.25 and 1.89, respectively. (c) Pdfs of 
the normalized increments of the series in Figure 1a. The plus signs correspond to $r = 2^{-14}$, while the circles 
refer to a distance $r = 2^{-6}$. The solid curve designates the Gaussian distribution for reference.}
\end{figure}

\newpage

\begin{figure}[ht]
\includegraphics[width=15pc]{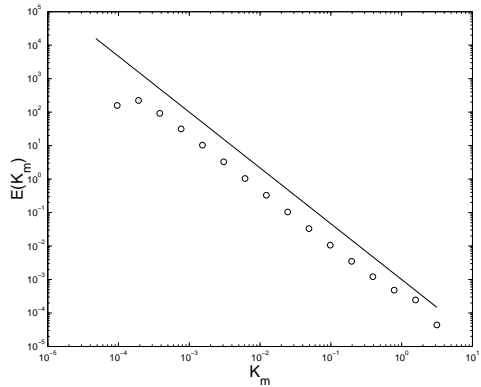}
\caption{Wavelet power spectrum (cirlces) of the series in Figure 1a. The -5/3 power law is also shown for comparison.}
\end{figure}

\newpage

\begin{figure}[ht]
\includegraphics[width=15pc]{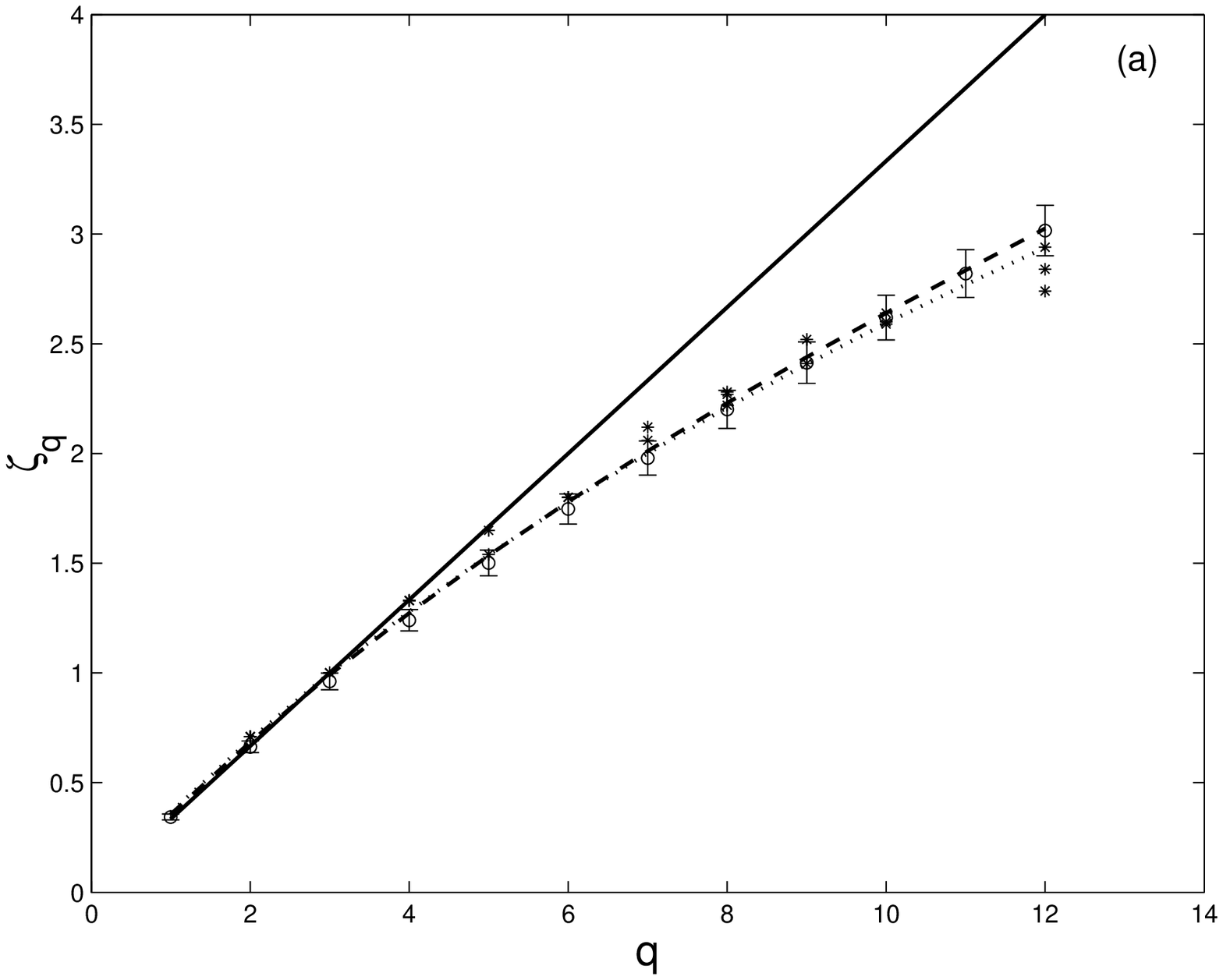}
\includegraphics[width=15pc]{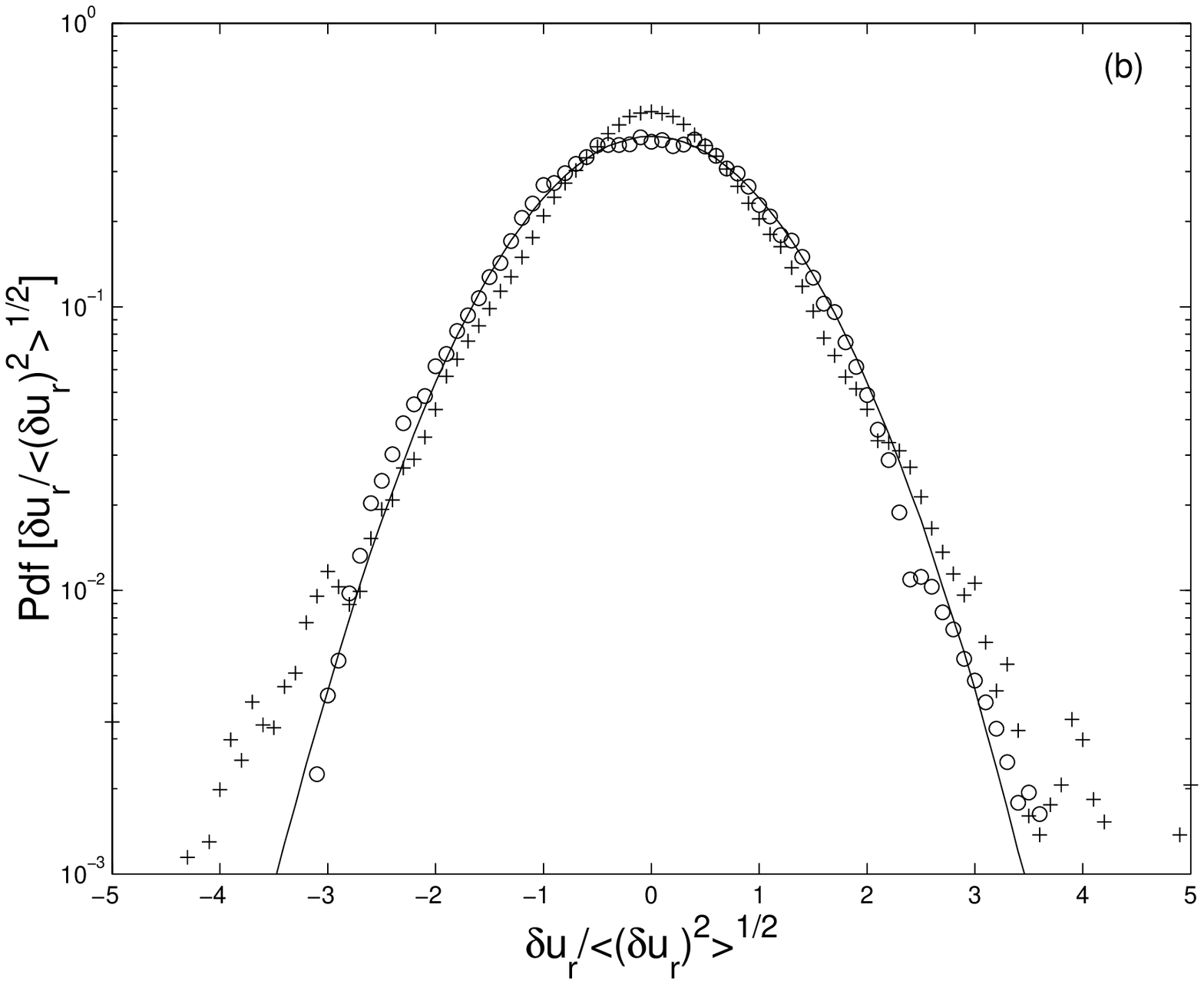}
\caption{(a) The scaling exponent function $\zeta_q$. The continuous, dashed and dotted lines denote the K41, 
Equation \ref{eq6}, and the She-L\'{e}v\^{e}que model predictions respectively. The circles with error bars (one standard 
deviation) 
are estimated values over one hundred realizations using $d_1 = -0.887$ and $d_2 = 0.676$. Experimental data of 
Anselmet et al.'s [5] is also shown for reference (star signs). (b) Pdfs of the normalized increments of the 
multiaffine series. The plus signs denote $r = 2^{-14}$, while the circles refer to a distance $r = 2^{-6}$. The solid 
curve designates the Gaussian distribution for reference.}
\end{figure}

\newpage

\begin{figure}[ht]
\includegraphics[width=17pc]{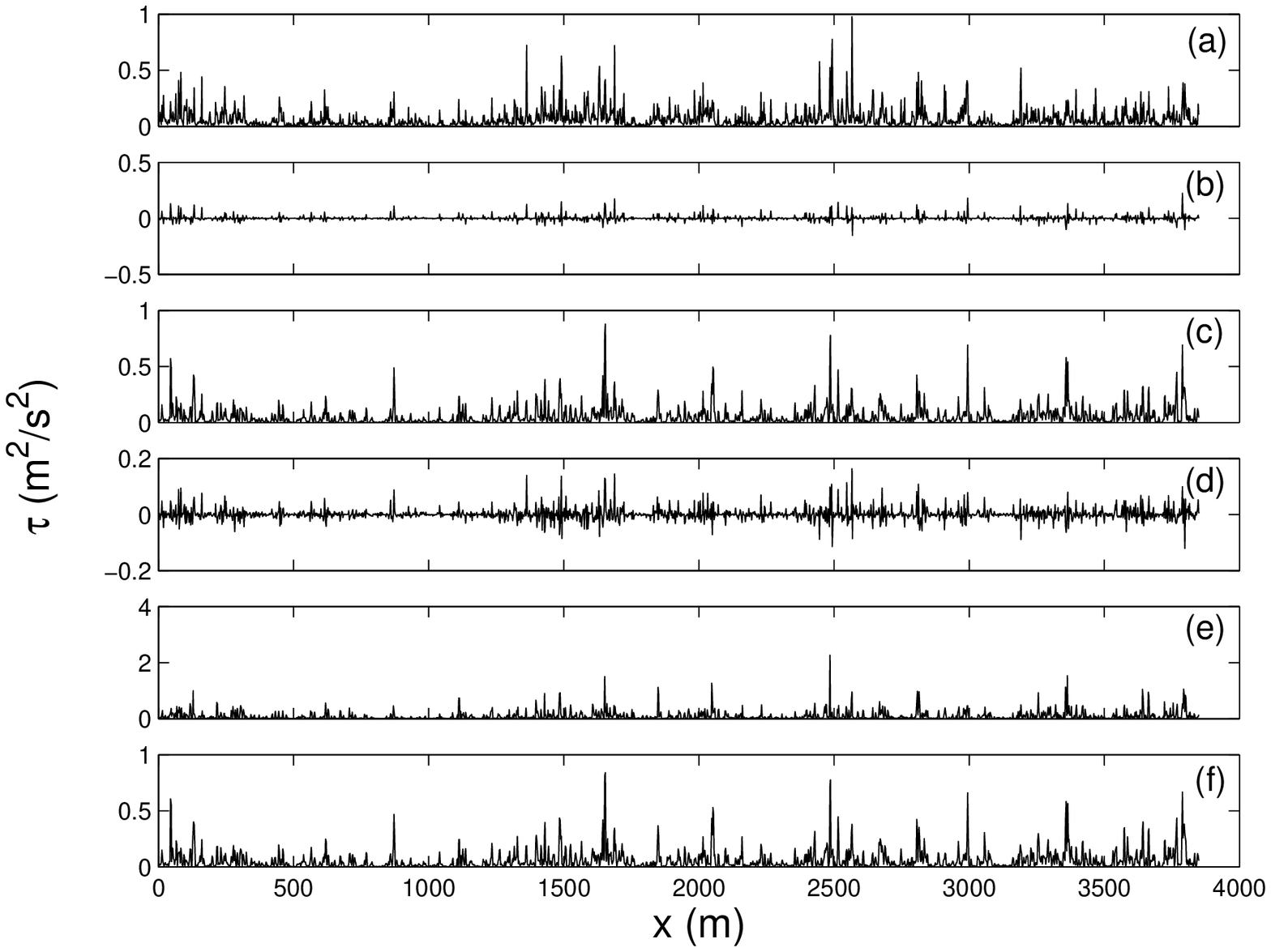}
\caption{A comparison of the real and modeled SGS stresses, computed from atmospheric boundary layer measurements, 
using 1-D filtering and Taylor's hypothesis. The filter width $\Delta$ is 2m. (a) Real, (b) Smagorinsky model, 
(c) Similarity model, (d) SGS kinetic energy based model, (e) Fractal model, and (f) Multiaffine model (M3).}
\end{figure}

\begin{figure}[ht]
\includegraphics[width=17pc]{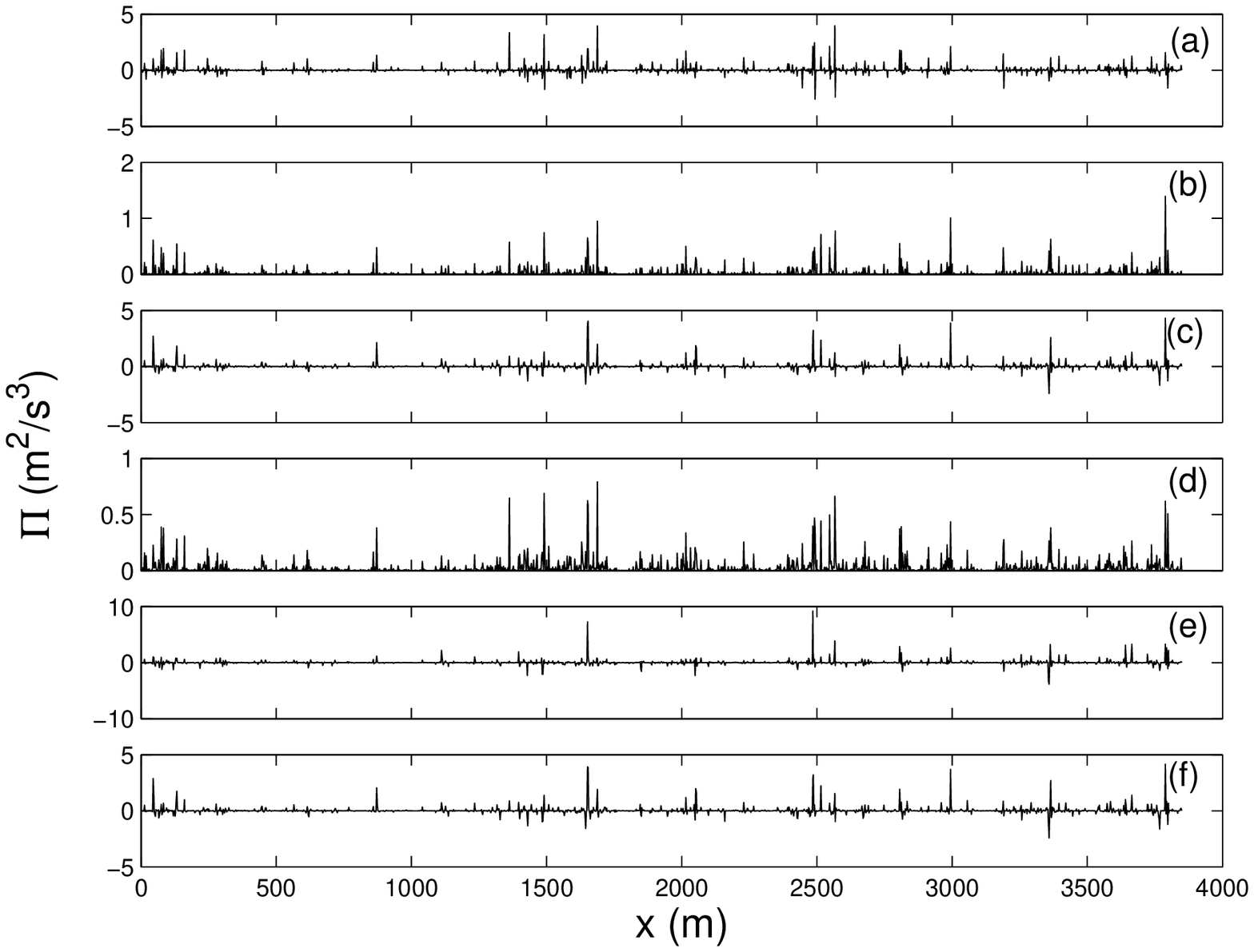}
\caption{A comparison of the real and modeled SGS energy dissipation rates, computed from atmospheric boundary layer 
measurements, 
using 1-D filtering and Taylor's hypothesis. The filter width $\Delta$ is 2m. (a) Real, (b) Smagorinsky model, 
(c) Similarity model, (d) SGS kinetic energy based model, (e) Fractal model, and (f) Multiaffine model (M3).}
\end{figure}

\newpage

\begin{figure}[ht]
\includegraphics[width=15pc]{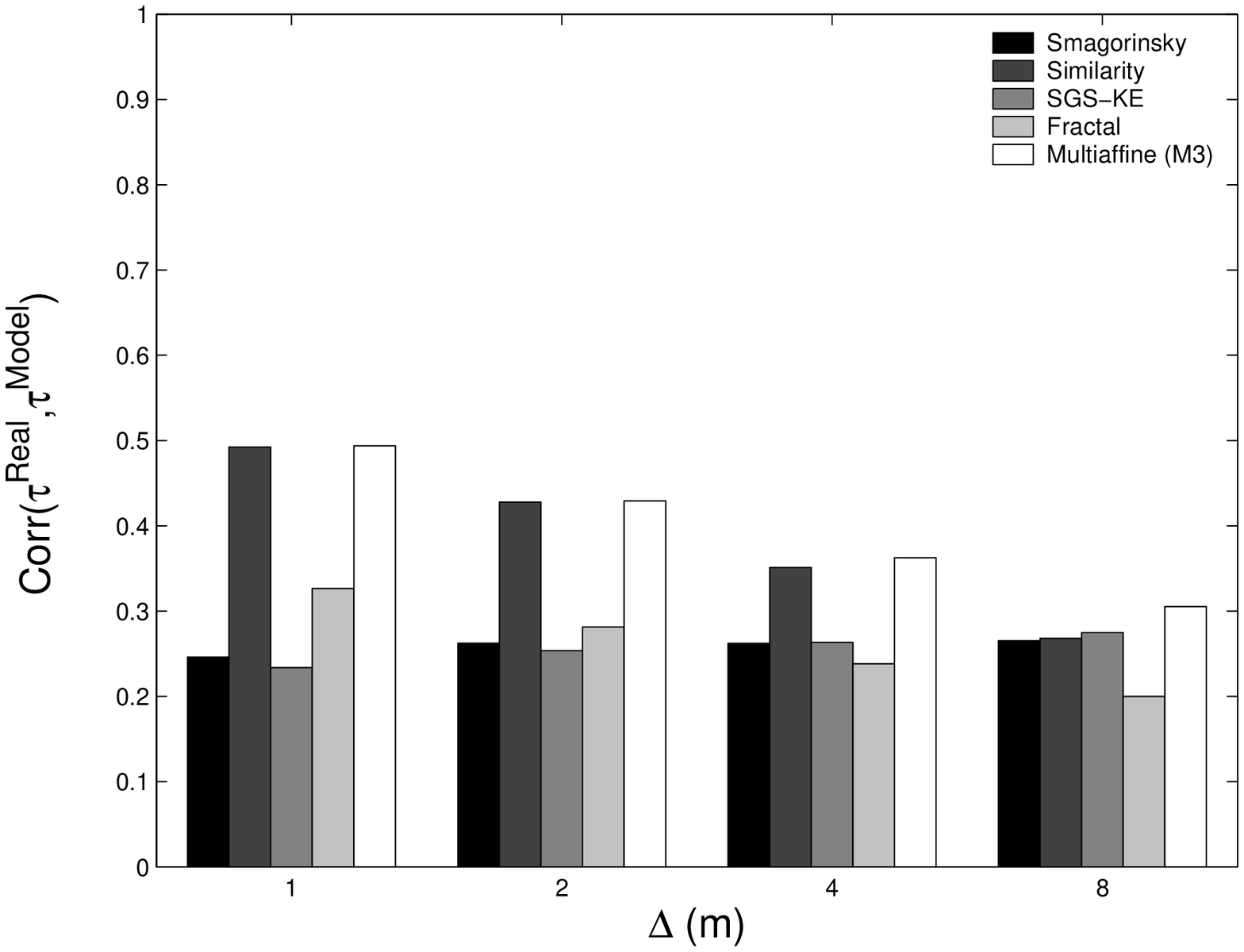}
\includegraphics[width=15pc]{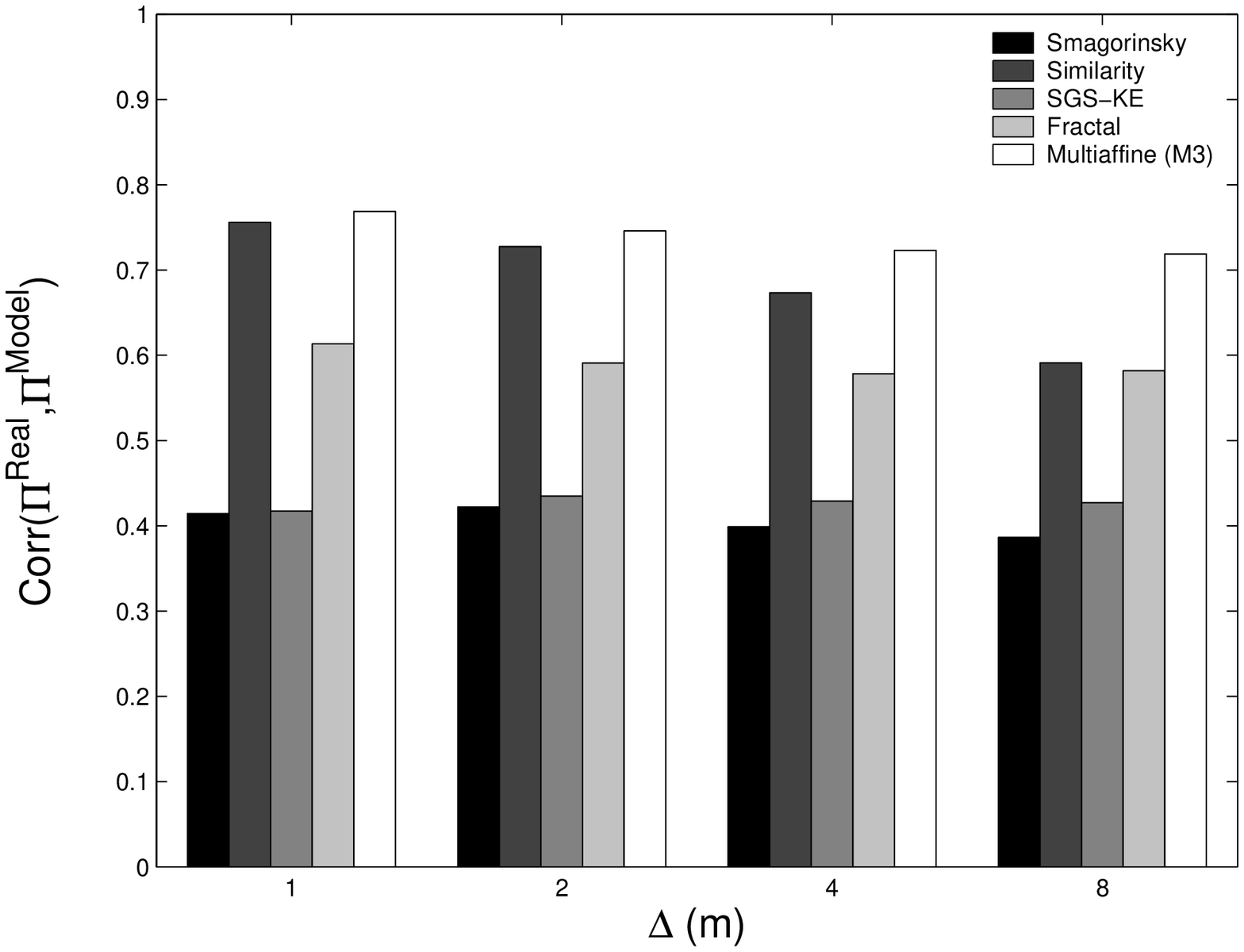}
\caption{(a) Correlation between observed and modeled subgrid-scale stresses and (b) correlation between observed and 
modeled subgrid-scale energy dissipations as a function of filter width $\Delta$. The results are based on 358 ABL 
turbulent velocity series measured during several field campaigns.}
\end{figure}

\newpage

\begin{figure}[ht]
\includegraphics[width=15pc]{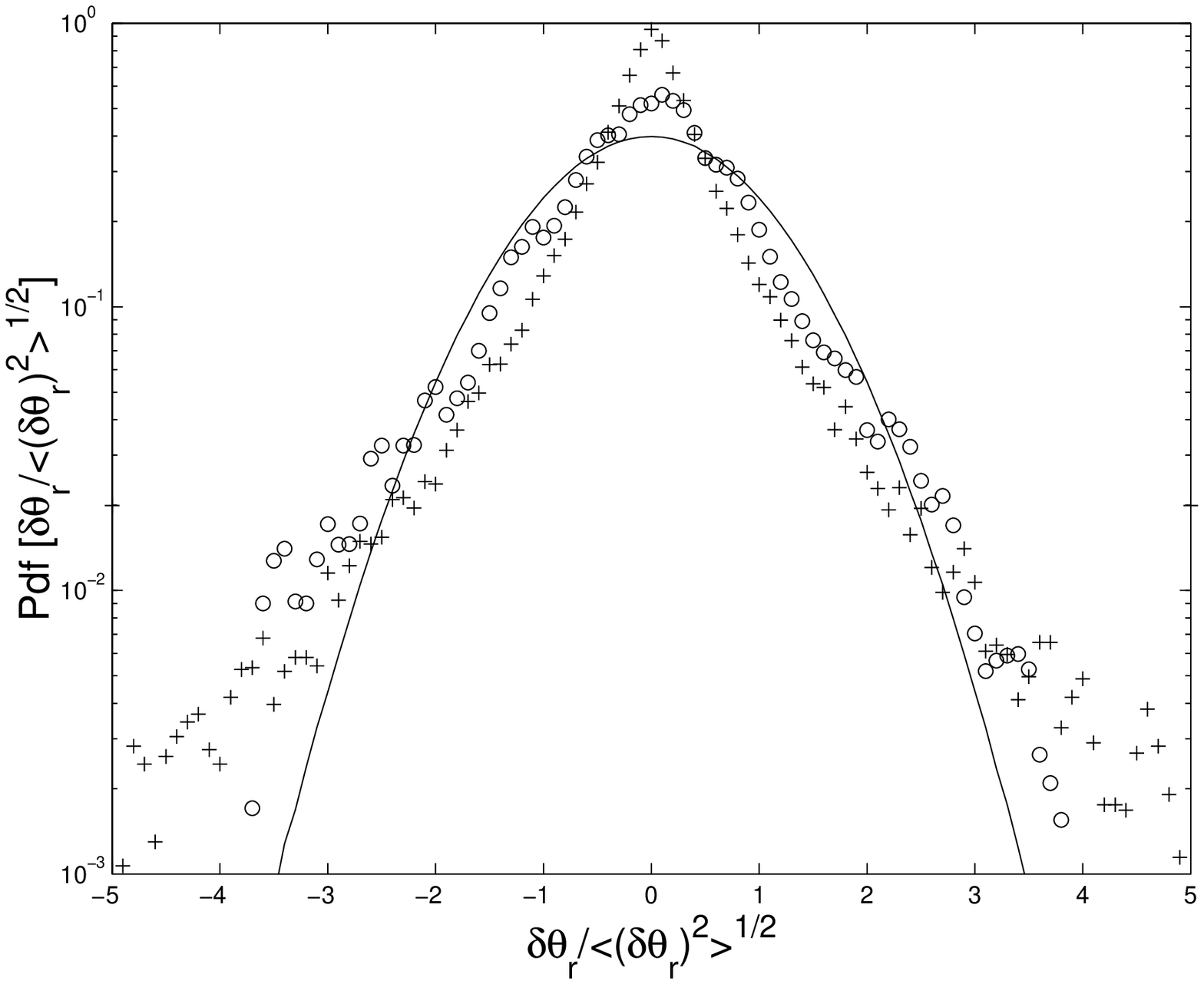}
\includegraphics[width=15pc]{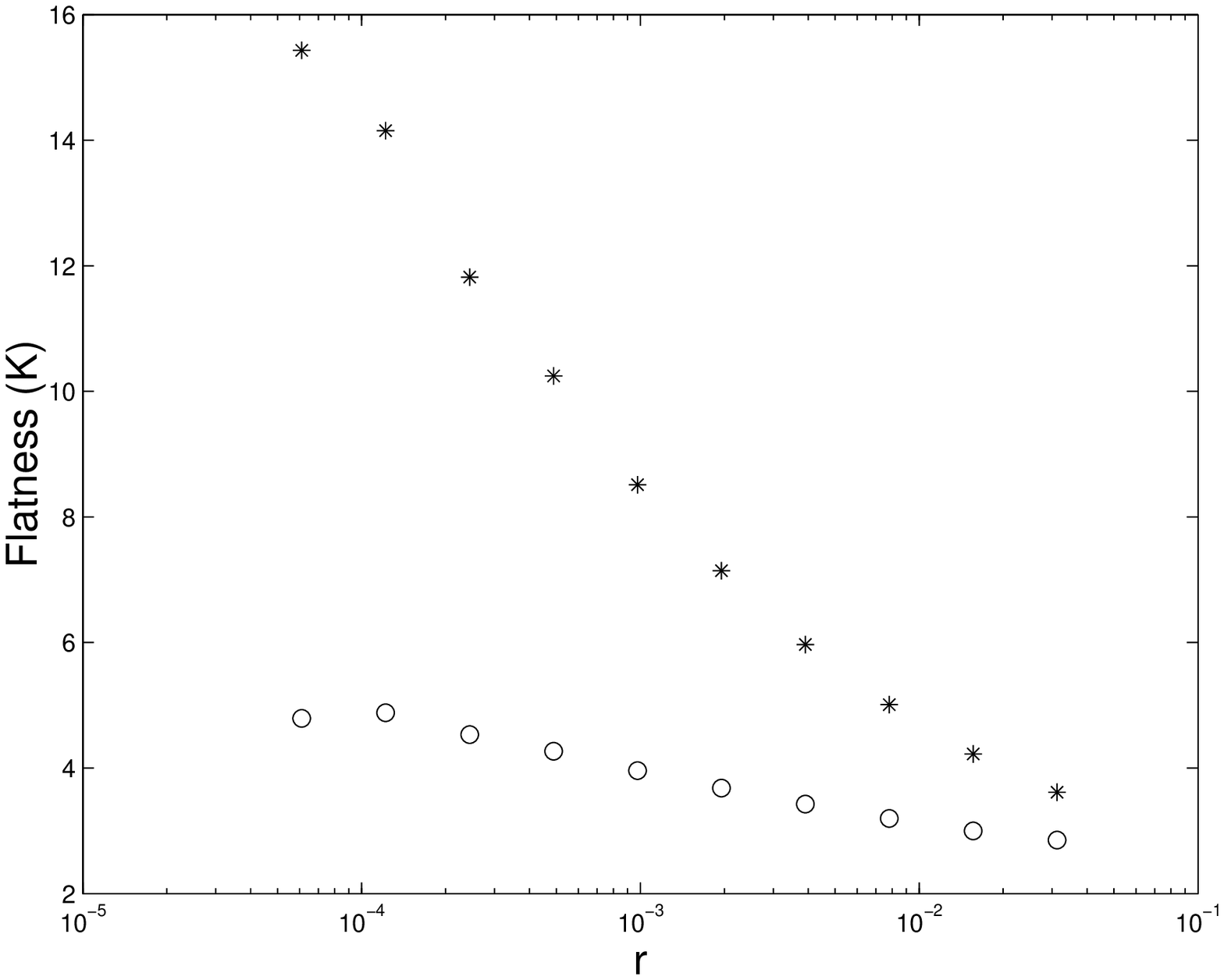}
\caption{(a) Pdfs of the normalized increments of the passive scalar Multiaffine series. The plus signs 
refer to distance $r = 2^{-14}$, while the circles to a distance $r = 2^{-6}$. The solid curve designates the Gaussian 
distribution for reference. 
(b) The flatness factors of the pdfs of the increments of the velocity (circles) and passive-scalar field (stars) 
as a function of distance $r$. Note that both the fields approach the Gaussian value of 3 only at large separation 
distances. Clearly the passive-scalar field is more non-Gaussian than the velocity field.}
\end{figure}


\begin{thebibliography}{}
\bibitem{Footnote1} In most of today's LES computations,
the filtering operation is basically implicit. This means that the process of numerical discretization (of grid size $\Delta$)
in itself creates the filtering effect. Although explicit filtering of the grid-resolved velocity field is desirable
from a theoretical point of view, limited computational resources make it impractical \cite{Eyink}. However, 
in a-priori studies of DNS, experimental or field observations, where fine resolution velocity fields are readily
available, the explicit filtering (typically with a Top Hat, Gaussian or Cutoff filter) followed by downsampling are 
usually performed to compute the resolved (and filtered) velocity $\tilde {u}_i$. 

This issue of filtering of velocity fields should not be confused with the explicit filterings done in the 
cases of Similarity, Dynamic, Fractal or Multiaffine models to compute the SGS stresses or other necessary SGS tensors.
\bibitem{Footnote2} {Following, \cite{Liu}, this filtering followed by downsampling 
  methodology is known in the literature as ``consistent'' a-priori analysis. }
\end{thebibliography}
\end{document}